\documentclass[preprint,tightenlines,showpacs,preprintnumbers,amsmath,amssymb,prb]{revtex4}

\usepackage{graphicx}
\usepackage{bm}

\begin{document}
\draft
\title{Geometrical and electronic structures of the (5, 3) single-walled gold
nanotube from first-principles calculations}

\author{Xiaoping Yang}

\affiliation{Group of Computational Condensed Matter Physics,
National Laboratory of Solid State Microstructures and Department
of Physics, Nanjing University, Nanjing 210093, P. R. China}
\affiliation{Department of Physics, Huainan Normal University,
Huainan, Anhui 232001, P. R. China}

\author{Jinming Dong}
\affiliation{Group of Computational Condensed Matter Physics,
National Laboratory of Solid State Microstructures and Department
of Physics, Nanjing University, Nanjing 210093, P. R. China}

\begin{abstract}
The geometrical and electronic structures of the 4 {\AA} diameter
perfect and deformed (5, 3) single-walled gold nanotube (SWGT)
have been studied based upon the density-functional theory in the
local-density approximation (LDA). The calculated relaxed
geometries show clearly significant deviations from those of the
ideally rolled triangular gold sheet. It is found that the
different strains have different effects on the electronic
structures and density of states of the SWGTs. And the small shear
strain can reduce the binding energy per gold atom of the deformed
SWGT, which is consistent with the experimentally observed result.
Finally, we found the finite SWGT can show the metal-semiconductor
transition.\\
\end{abstract}


\maketitle

Nanowires and nanotubes have raised extensive interest, which is
motivated both by their special electrical and mechanical
properties as well as by their potential applications in future
nanostructure materials. For example, the unique electronic and
mechanical properties of carbon nanotube (CNT) [1-4] are proved to
be a rich source of new fundamental physics and also a promising
candidate of nanoscale wires, transistors and sensors. So, the
cylindrical tubes and nanowires made of other materials are also
the subject of intense researches. For example, Si, BN,
SiSe$_{2}$, WS$_{2}$, MoS$_{2}$, NiCl$_{2}$, and various metallic
nanowires [5], e.g., gold nanowires, are experimentally studied
[6].

The long gold nanowires with diameter less than 2 nm have been
experimentally synthesized in an ultra-high vacuum (UHV) ---
transmission electron microscope (TEM) with the electron-beam
thinning technique [7], showing a coaxial helical multishell (HMS)
structure, which is similar to that of the multiwalled CNT except
that the honeycomb network of carbon atoms in the CNT is replaced
by a triangular network of gold atoms (Fig. 1) [7-10]. The
triangular network can be deformed due to the shear strain
$\varepsilon $ caused by the metallic bond. A helical double
walled $n$-$n^{'}$ gold nanowire is composed of two coaxial tubes
with $n$ and $n^{' }$helical gold atomic rows in the outer and
inner tube, respectively, which coiled round the tube axis. It is
found that the inner and outer shells always have even-odd or
odd-even coupling, and the difference ($n-n^{'}$) between their
number of atomic row is equal to seven (a single atom chain is
regarded as "0"-membered in this discussion). Nanowires with $n
\le 6$ had been supposed to have only a single-walled shell [7],
called as SWGT, which has recently been synthesized experimentally
at 150 K in an UHV---TEM [11]. The thinnest SWGT, 4 {\AA} in
diameter, was found to be the (5, 3), composed of five atomic rows
coiled round the tube axis.

The geometrical structure of the SWGT is schematically shown in
Fig. 1(a), where
$\mathord{\buildrel{\lower3pt\hbox{$\scriptscriptstyle\rightharpoonup$}}\over
{a}} _1 $ and
$\mathord{\buildrel{\lower3pt\hbox{$\scriptscriptstyle\rightharpoonup$}}\over
{a}} _2 $ denote two basis vectors of the triangular network. The
triangular network is deformed by the shear strain $\varepsilon$
in
$\mathord{\buildrel{\lower3pt\hbox{$\scriptscriptstyle\rightharpoonup$}}\over
{a}} _2 $ direction. The tube unit cell is defined by the vector
$\mathord{\buildrel{\lower3pt\hbox{$\scriptscriptstyle\rightharpoonup$}}\over
{C}} (n,m) =
n\mathord{\buildrel{\lower3pt\hbox{$\scriptscriptstyle\rightharpoonup$}}\over
{a}} _1 + (m + n\varepsilon /
2)\mathord{\buildrel{\lower3pt\hbox{$\scriptscriptstyle\rightharpoonup$}}\over
{a}} _2$ and its orthogonal $\overrightarrow {OB} $ parallel to
the tube axis. As used in the single-walled carbon nanotube
(SWNT), here, a pair of integers, ($n$, $m$), is also used to
represent a SWGT consisting of $n$ close-packed atomic rows with a
helical angle ranging from 0\r{ } ($m=n/2$) to 30\r{ }($m=0$),
which is defined as the angle between the atom row and the tube
axis.

It is well known that a mechanical deformation of a SWNT affects
heavily its electronic structures [12-14]. So, it is natural to
ask what happens for the SWGT due to different deformations. In
this paper, we use the first-principles calculation to study the
structural and electronic properties of the 4 {\AA} diameter SWGT
(5, 3) exerted by three kinds of strains, i.e. the uniaxial
stretch, torsional and shear strains. In addition, we investigate
the effect of the finite length of SWGT on its electronic
structures.

The total energy plane-wave pseudopotential method [15] has been
used in our calculations within the framework of local density
approximation (LDA). The ion-electron interaction is modeled by
ultra-soft local pseudopotentials of the Vanderbilt form [16] with
a uniform energy cutoff of 224.7 eV. There are 21 uniformly
distributed \textit{k} points along the nanotube symmetry axis
\textit{$\Gamma $X}. The supercell geometry for the SWGT has been
used [17], in which the tubes are aligned in a hexagonal array
with the closest distance between the adjacent tubes being 20
{\AA}, larger enough to prevent the tube-tube interactions. Using
the same computational method, we have also calculated the band
structure of the SWGT (8, 4) plus one atomic row at the center of
the tube, and the obtained result is well consistent with that in
Ref. [8].

Firstly, we study the structural and electronic properties of the
fully relaxed perfect SWGT (5, 3) and a uniaxially stretched one
(with uniaxial strain of 0.05 and 0.1), for which the obtained
structure parameters are shown in Table I. From it we can find
that the diameter of the relaxed perfect (5, 3) tube is 0.2 {\AA}
larger than that (4.03 {\AA}) of ideal tube rolled up from the
gold triangular network sheet, while its lattice constant along
the tube axis is 1.137 {\AA} smaller than that (21.895 {\AA}) of
an ideal tube. This is due to the curvature effect that weakens
the bonds that are wrapping around the circumference of the tube.
So, it is also found that the average bond length along the axis
of relaxed perfect tube is shorter than those in the direction of
the circumference. However, under the uniaxial strain of 0.05 and
0.1, the diameter of the deformed tube gradually recovers that of
the ideal tube. In addition, the binding energy per gold atom for
the uniaxial stretched tube is obviously found to be higher than
that of the relaxed perfect tube.

The calculated electronic band structures and density of states
are shown in Fig. 2(a)-2(c) and 3(a)-3(b) (only the DOSs of the
uniaxial stretch strain of 0.1 are given as an example for the
case of the uniaxial stretch), respectively. Comparing Fig. 2(a)
with 2(b) and 2(c), we can find that the uniaxial stretch strain
makes the bands near the Fermi level move toward the higher
energy, but keeping always five conduction channels on the Fermi
level. So, the SWGT (5, 3) is not affected sensitively by the
uniaxial strain. And the shifts of the energy bands lead to the
shifts of the DOS peaks, which can be seen clearly in Fig. 3(a)
and (b). Three DOS peaks at -1.74, -0.15 and 1.76 eV (Fig. 3(a))
are blue-shifted to -1.71, 0.03 and 1.86 eV under uniaxial strain
of 0.1, respectively (Fig. 3(b)). At the same time, the DOS peak
at -1.92 eV (Fig. 3(a)) in valence band is red-shifted to -2.01 eV
(Fig. 3(b)). The uniaxial stretch of 0.1 leads also to
disappearance of the DOS peak at -1.32 eV (Fig. 3(a)) in Fig.
3(b).

Secondly, we have studied the geometrical structures and
electronic properties of the torsionally deformed SWGT (5, 3),
whose ball-and-stick pictures are also shown in Fig. 1. When the
torsional strain takes positive +4.302\r{ } or negative -6.562\r{
}, it means the strain is applied along CB (OA) or BC (AO)
direction in Fig. 1(a). The obtained structural parameters of the
fully relaxed tubes are shown in Table I too, from which it is
found that the diameters of the both deformed (5, 3) tubes are
larger than that of the perfect one. The positive torsion makes
the strained bond $b$ (and $a)$ longer (shorter) than that without
the strain; while the negative torsion leads to a reverse case.
Meanwhile, the strained bond $c$ always becomes shorter than that
before with different torsional strains. The binding energies per
gold atom under the positive and negative torsions are larger than
that of perfect tube by maximal 0.031 eV.

The calculated electronic band structures and density of states are shown in
Fig. 2(d), 2(e) and Fig. 3(c) (choose only the torsional strain of +4.302\r{
} as an example). Since the total gold atom numbers are 23 and 5 per unit
cell under the torsional strains of +4.302\r{ } and -6.562\r{ },
respectively, some energy bands clearly disappear in Fig. 2(d) and 2(e),
especially for the negative -6.562\r{ }. But they still keep five conduction
channels on the Fermi level, which clearly indicates the torsional strain
does not affect more heavily the SWGT (5, 3) than the uniaxial one. The
complex change of the energy bands makes the shift of DOS peaks different
from that under the uniaxial stretch strain. Under the torsional strain of
+4.302\r{ }, the three DOS peaks at -1.92, -1.32 and -0.15 eV in Fig. 3(a)
are red-shifted to -1.93, -1.48 and -0.22 eV, respectively in Fig. 3(c), and
DOS peaks at -1.74 and 1.76 eV in Fig. 3(a) are blue-shifted to -1.71 and
1.85 eV in Fig. 3(c), respectively.

Finally, we study the structural and electronic properties of
deformed SWGT (5, 3) under shear strain. Due to limitation of the
computational time and cost, we can only select the shear strain
of $\varepsilon = 0.0314$ along the basis vector
$\mathord{\buildrel{\lower3pt\hbox{$\scriptscriptstyle\rightharpoonup$}}\over
{a}} _2 $ (Fig. 1(a)), making the one unit cell of the deformed
tube contain 33 gold atoms, which is schematically shown in Fig.
1. The obtained fully relaxed structure parameters are shown in
Table I too, and from it we can find that the diameter of the
shear-strain deformed SWGT (5, 3) is only slightly larger than
that of the perfect one. But its binding energy per gold atom is
smaller than that of perfect tube by 0.008 eV, which means it is
more stable than the perfect tube, being consistent with the
experimentally observed result [11]. Its calculated electronic
band structures and DOSs are shown in Fig. 2(f) and Fig. 3(d).
Comparing Fig. 2(f) with Fig. 2(a), we can find the shear strain
of 0.0314 causes less change of the band structure since its total
atom number per unit cell is 33, only 5 less than that (38) of the
perfect tube, and only a few bands disappear. So, the conduction
channel on the Fermi level still remains five. As for its DOSs, we
found three DOS peaks at -1.92, -1.32 and -0.15 eV in Fig. 3(a)
are slightly red-shifted to -1.93, -1.36 and -0.17 eV in Fig.
3(d), respectively, and only the DOS peak at 1.76 eV in Fig. 3(a)
is slightly blue-shifted to 1.78 eV in Fig. 3(d). Meanwhile, the
peak position at -1.74 eV in Fig. 3(a) is not changed, but its
height is decreased. And a new weaker DOS peak emerges at -1.55 eV
in Fig. 3(d), which corresponds to the shoulder between the DOS
peak values -1.74 and -1.32 eV in Fig. 3(a). Experimentally, the
synthesized SWGT (5, 3) has an intrinsic shear strain $\varepsilon
$ of 0.005, which is far smaller than that used in our
calculation. So, based on our result the shear strain effect on
the electronic structure of the SWGTs should be very small. In
addition, we found that the deformed (5, 3) tubes under the
torsional strain of 1.006\r{ } and shear strain of 0.0314 have the
very similar structural parameters (see Table 1) and the
electronic structures. And the difference between their DOS peak
positions is also small, lying in a range of about $\pm 0.04$ eV,
which indicates that the very small shear strain plays an
equivalent role in the SWGT (5, 3) as a torsional strain.

In addition, it is noted from Fig. 3 that the DOS peaks in valence
bands lower than -0.5 eV are mainly contributed by the $d$-orbital
electrons; while the $s$-orbital electrons dominate the energy
bands higher than -0.5 eV, which is similar for the deformed tubes
under three different kinds of strains.

In addition to the SWGT (5, 3), we also study the electronic
properties of the deformed SWGT (5, 0) (helical angle is 30\r{ })
and SWGT (6, 3) (helical angle is 0\r{ }) under the similar
uniaxial stretch of 0.1 and torsional strains (+4.715\r{ } for (5,
0) and +6.587\r{ } for (6, 3)). The band structures of (5, 0) and
(6, 3) are shown in Fig. 4. It can be seen in Fig. 4 that no
significant changes are found, and the Fermi level in every panel
is always crossed by five conduction bands (the two-degenerate
state is labeled by the arrow). Under the uniaxial strain, the
energy bands near the Fermi level move toward the higher energy
area, and the torsional strain eliminate the double-degeneracy of
the energy band near the Fermi level. The different changes of the
energy bands under the two kinds of the strains make inevitably
the shift of the DOS peaks different, which is consistent with the
above results of SWGT (5, 3). It is found that the SWGT (6, 3) has
the same five conduction channels as (5, 3) and (5, 0) tubes
indicating clearly no direct correlation between the numbers of
the atom rows and the conduction channels.

It is known that most of the deformed SWNTs show a
metal-semiconductor transition (MST), occurring periodically with
increasing strain [18-20], but why do the deformed SWGTs always
remain metal? There are two inequivalent carbon atoms in one unit
cell of the hexagonal graphite sheet, causing both of bonding and
anti-bonding bands, which cross at six points, and so are easy to
be separated by applied strains. In contrast, only one gold atom
exists in one unit cell of the triangular gold sheet, leading to
only one energy band. So, the uniaxial and torsional strain can
not make the periodic infinite SWGT show the MST, which is already
proved by the above calculated results. However, the practical
SWGTs have the finite length of usual several nanometers (Ref.
11), and so it is necessary to study the finite length effect on
the physical properties of SWGT. Our numerical calculation
indicates that the SWGT (5, 3) with a length of 4.15 nm (about two
times larger than the perfect tube period) shows the
characteristic of the semiconductor with its gap of 0.11 eV. And
it is found the energy gap will oscillate with increasing length,
which is similar to that of SWNT [21-22]. Due to limitation of
computation time, we take only the results of SWGT (5, 0) as an
example in Table II, from which we know the SWGT (5, 0) with the
optimized length of about 1.37, 1.83, 2.29, 3.2 and 3.67 nm have
the energy gap of 0.75, 0.3, 0.16, 0.23 and 0.165 eV,
respectively, but the ones with the optimized length of 2.75 and
4.12 nm remain metal. Therefore, by applied strains, the
finite-length SWGT can also show the similar MST as the SWNT.

In summary, we have systematically studied the structural and
electronic properties of the perfect and deformed SWGT (5, 3),
under three kinds of strains, i.e. the uniaxial stretch, torsional
and shear strains. It is found that there is no direct correlation
between number of the atom rows forming the SWGT and that of
conduction channels, and it is difficult to change the conduction
channel number under the applied strains, which always keeps five.
The small shear strain can reduce the binding energy per gold atom
of SWGT, which is consistent with the experimentally observed
result. Compared with the SWNT, the geometrical and electronic
structures of the SWGT are more stable under the applied strains,
making absent the MST for the deformed infinite SWGT, which
contrasts to those in SWNT. However, the SWGT with the finite
length is found to show the characteristic of the semiconductor.
We hope the obtained results in this paper will be helpful to
design nanoscale electronic device based on the SWGT.

This work was supported by the Natural Science Foundation of China
under Grant No. 10474035, No. A040108, and also support from a
Grant for State Key program of China through Grant No.
2004CB619004.

\newpage

\begin{center}
\textbf{TABLE}
\end{center}

\begin{table}[htbp]
\textbf{Table I.}Calculated structural parameters of the fully
relaxed perfect and deformed SWGT (5, 3) in a 20 {\AA}$\times $20
{\AA}$\times $C$_{l}$ {\AA} hexagonal supercell. The parameters
(a, b, c) are defined as in Fig. 1(a). And $E_{b}$ is the binding
energy per gold atom.
\begin{tabular}
{|p{90pt}|p{63pt}|p{81pt}|p{45pt}|p{45pt}|p{45pt}|p{54pt}|} \hline
\textit{(5,3) tube}& \textit{Diameter (}\textit{{\AA}}) \par
\textit{(relaxed)}& \textit{lattice constant }$C_{l}$ \par
\textit{(relaxed)}& $a$ \textit{({\AA})}& $b$ \textit{({\AA})}&
$c$ \textit{({\AA})}&
$E_{b}$\textit{ (eV)} \par \textit{per atom} \\
\hline Perfect& 4.23& 20.758& 2.835& 2.819& 2.754&
-2.598 \\
\hline Stretched 0.05& 4.15& 21.796& 2.848& 2.794& 2.889&
-2.565 \\
\hline Stretched 0.1& 4.04& 22.834& 2.851& 2.754& 3.024&
-2.501 \\
\hline Torsion +1.006\r{ }& 4.27& 17.825& 2.821& 2.853& 2.731&
-2.606 \\
\hline Torsion +4.302\r{ }& 4.32& 12.327& 2.762& 2.930& 2.744&
-2.583 \\
\hline Torsion -6.562\r{ }& 4.31& 2.709& 3.010& 2.756& 2.709&
-2.567 \\
\hline Shear 0.0314& 4.27& 17.826& 2.821& 2.853& 2.731&
-2.606 \\
\hline
\end{tabular}
\label{tab1}
\end{table}

\begin{table}[htbp]
\textbf{Table II.}Energy gap versus tube length for the finite
SWGT (5, 0).
\begin{tabular}
{|p{113pt}|p{44pt}|p{44pt}|p{44pt}|p{44pt}|p{44pt}|p{44pt}|p{44pt}|}
\hline The original length of tube (nm)& 1.5& 2.0& 2.5& 3.0& 3.5&
4.0&
4.5 \\
\hline The optimized length of tube (nm)& 1.37& 1.83& 2.29& 2.75&
3.2& 3.67&
4.12 \\
\hline Semidconductor (S) \par or Meal (M)& S& S& S& M& S& S&
M \\
\hline Energy gap (eV)& 0.75& 0.3& 0.16& 0.0& 0.23& 0.165&
0.0 \\
\hline
\end{tabular}
\label{tab1}
\end{table}

\newpage

\begin{figure}[htbp]
\includegraphics[width=0.8\columnwidth]{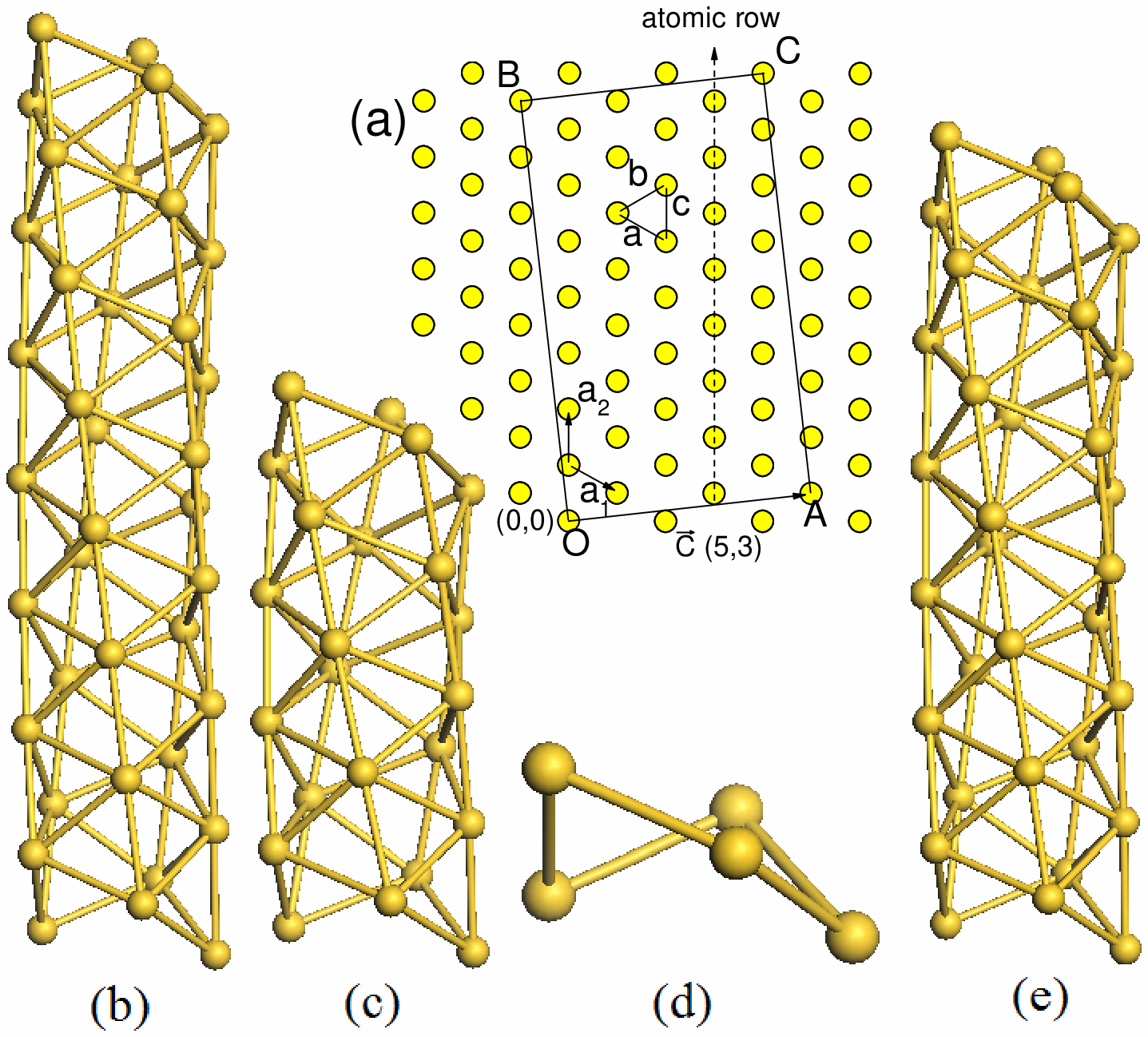}
\label{fig1} \caption{(Color online) (a) The triangular network
sheet of gold atoms forming the SWGT. The nearest neighbor distance
between two gold atoms is chosen to be 2.9 {\AA}. Ball-and-stick
model of one unit cell for the 4 {\AA} diameter SWGT (5, 3): (b)
undeformed structure with its total atom number of 38; (c) torsional
strain of +4.302\r{ } (23 atoms); (d) torsional strain of -6.562\r{
} (5 atoms); (e) shear strain of 0.0314 (33 atoms).}
\end{figure}

\begin{figure}[htbp]
\includegraphics[width=0.8\columnwidth]{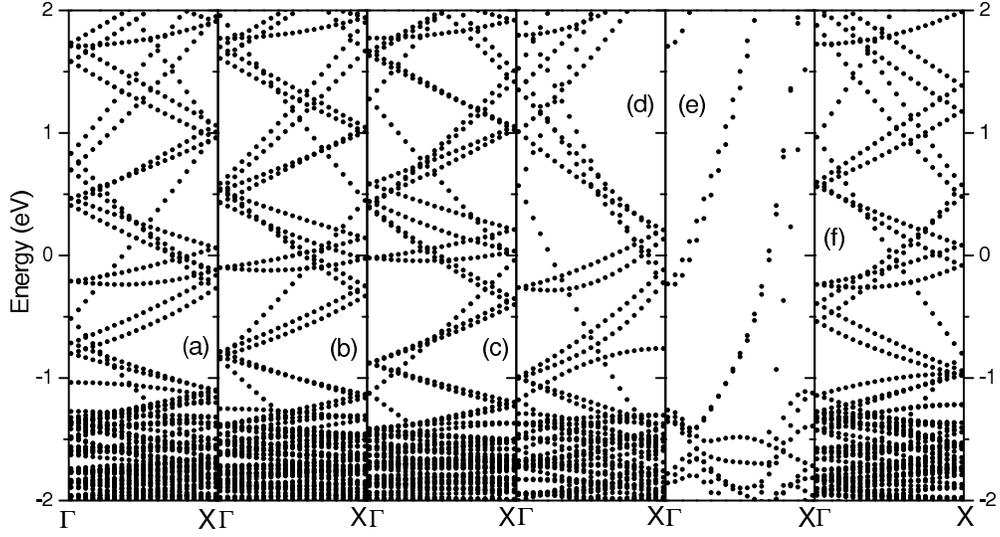}
\label{fig2} \caption{Calculated band structures of : (a) perfect
SWGT (5, 3); (b), (c), (d), (e) and (f) deformed ones, respectively,
under uniaxial stretch strain of 0.05 and 0.1, torsional strain of
+4.302\r{ } and -6.562\r{ }, and shear strain of 0.0314. The Fermi
level is set at zero.}
\end{figure}

\begin{figure}[htbp]
\includegraphics[width=0.8\columnwidth]{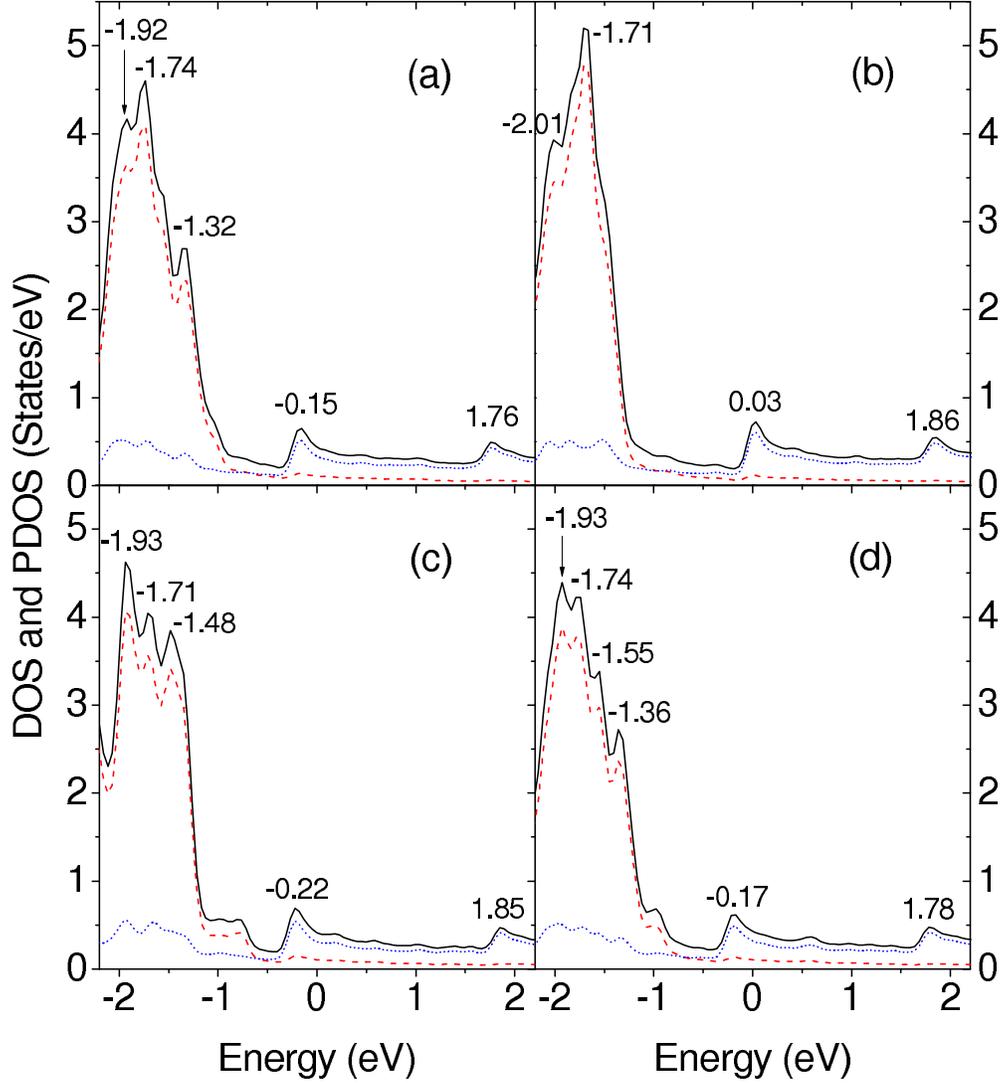}
\label{fig3} \caption{(Color online) Calculated total DOSs (solid
line), PDOSs for $d$ electrons (dash line) and $s$ electrons (dot
line): (a) perfect SWGT (5, 3), (b), (c) and (d) deformed ones,
respectively, under uniaxial strain of 0.1, torsional strain of
+4.302\r{ }, and shear strain of 0.0314. The Fermi level is set at
zero.}
\end{figure}

\begin{figure}[htbp]
\includegraphics[width=0.8\columnwidth]{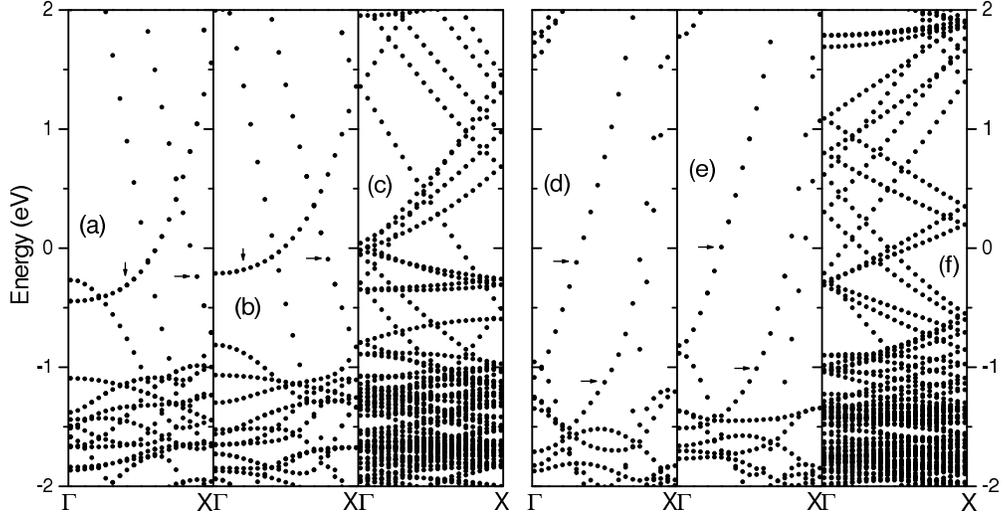}
\label{fig4} \caption{Calculated band structures of : (a) perfect
SWGT (5, 0) (10 atoms); (b) and (c) deformed ones, respectively,
under uniaxial stretch strain of 0.1, torsional strain of +4.715\r{
} (35 atoms); (d) perfect SWGT (6, 3) SWGT (6 atoms); (e) and (f)
deformed ones, respectively, under uniaxial stretch strain of 0.1,
torsional strain of +6.587\r{ } (45 atoms). The Fermi level is set
at zero.}
\end{figure}
\end{document}